# Brain Connectivity Analysis Methods for Better Understanding of Coupling

Revati Shriram[1,2]

[1]Research Scholar, Sathyabama University, Chennai.

[2]Cummins College of Engg for Women, Pune, INDIA

revatishriram@yahoo.com

Dr. M. Sundhararajan

Shri Lakshmi Ammal Engg. College, Chennai, INDIA

msrajan69@gmail.com

Nivedita Daimiwal

Cummins College of Engineering for Women, Pune, INDIA

nivedita.daimiwal@gmail.com

*Abstract—* **Action, cognition, emotion and perception can be mapped in the brain by using set of techniques. Translating unimodal concepts from one modality to another is an important step towards understanding the neural mechanisms. This paper provides a comprehensive survey of multimodal analysis of brain signals such as fMRI, EEG, MEG, NIRS and motivations, assumptions and pitfalls associated with it. All these non-invasive brain modalities complement and restrain each other and hence improve our understating of functional and neuronal organization. By combining the various modalities together, we can exploit the strengths and flaws of individual brain imaging methods. Integrated anatomical analysis and functional measurements of human brain offer a powerful paradigm for the brain mapping. Here we provide the brief review on non invasive brain modalities, describe the future of co-analysis of these brain signals.**

*Keywords- EEG, fMRI, MEG, NIRS and BMI.*

## I. INTRODUCTION

The in vivo measurement of blood perfusion in an organ has been a topic of interest for many years. Modern imaging methods provide the opportunity for non-invasive *in vivo* study of human organs and can provide measurements of local neuronal activity of the living human brain *(A Toga et al, 2001)*. These imaging modalities can be divided into two global categories: Functional Imaging or Structural Imaging *(Fantini et al, 2001)*. Functional imaging technique can be used along with the structural imaging to better examine the anatomy and functioning of particular areas of the brain in an individual.

### Functional Imaging:

Functional imaging represents a range of measurement techniques in which the aim is to extract quantitative information about physiological function from image-based data. The emphasis is on the extraction of physiological parameters rather than the visual interpretation of the images. Functional modalities include Single Positron Emission Computed Tomography (SPECT) and Positron Emission Tomography (PET), these are the nuclear medicine imaging modalities. Along with them Functional Magnetic Resonance Imaging (fMRI), Electroencephalogram (EEG), Magnetoencephalogram (MEG), Electrical Impedance Tomography (EIT) can also be named as a functional imaging techniques *(Fantini et al, 2001)*.

### Structural Imaging:

Structural imaging represents a range of measurement techniques which can display anatomical information. These modalities include X-ray, Computer Tomography (CT), Magnetic Resonance Imaging (MRI), Transcranial Magnetic Stimulation (TCM) and Ultrasound (US) *(Fantini et al, 2001)*. There are many reasons to determine the regional blood flow in organs such as in the brain or kidney, or in cancerous tissue regions of the body. For example, the assessment of cerebral blood flow and its autoregulation can be used to investigate the normal physiology and the nature of various diseases of the brain *(T. S. Koh, Z. Hou, 2002)*. Also, the efficacy of radiotherapy treatment of cancer cells depends on the local oxygen concentration which is governed by the local blood flow. A convenient, minimally invasive method of assessing blood flow within organs is hence constantly being sought *(A Toga et al, 2001)*.

## II. NEUROIMAGING METHODS

### EEG: Electroencephalogram

EEG signal originates mainly in the outer layer of the brain mainly known as the cerebral cortex, a 4–5mm thick highly folded brain region responsible for activities such as movement initiation, conscious awareness of sensation, language, and higher-order cognitive functions (*E.B.J. Coffey et al, 2010*). EEG signal describes electrical activity of the brain measured by unpolarized electrodes and belongs to the group of stochastic (random) signals in frequency band of about 0 – 50 Hz with rather high time resolution (units - tens of ms) (*T. Heinonen et al, 1999*). In contrast, the anatomical localization of specific sources of the electrical activity is very imprecise. Electrical impulses, which come from deep centers of the brain, are not possible to measure directly using the scalp EEG approach (*R. Labounek el at, 2012*)

### fMRI: Functional Magnetic Resonance Imaging

fMRI is a method of brain activity exploration based on repeated brain volume scanning by a MRI tomography (*E.B.J. Coffey et al, 2010, R. Misri et al, 2011*). The measured local signal corresponds to changes in the ratio of paramagnetic deoxyHb (HHb) and diamagnetic oxyHb (HbO$_2$). It is denoted as a BOLD signal (Blood Oxygenation Level Dependent) (*A. Buchweitz et al,2009*). 3D results, which are obtained from fMRI, have an excellent spatial resolution, while its time resolution is significantly worse than for EEG because the period of one brain scan is in the order of seconds. (*R. Labounek el at, 2012*)

### NIRS: Near Infrared Spectroscopy

Near Infrared (IR) light (wavelength 600 - 1000 nm) easily penetrates the biological tissue (*F. Irani et al, 2007, M. Tamura et al, 1997*). NIRS is based on the observation that the properties of light passing through a living tissue are influenced by the functional state of the tissue. It is a non-invasive method to measure oxygenation in a localized tissue and measures the transmission of infrared light through biological tissue (*G. Strangman et al, 2005*). This indicates changes in oxygenation and the concentration of tissue chromophores such as total haemoglobin concentration (tHb) with its constituent oxygenated haemoglobin (HbO$_2$) and deoxygenated haemoglobin (HHb) and cytochrome oxidase (CytOx) (*Nagdyman et al., 2003*). NIRS signal obtained is based on capillary-oxygenation-level-dependent (COLD) signal. Figure 1 shows the light propagation path inside the skull and absorption spectra of HHb, HbO$_2$, water and CytOx.

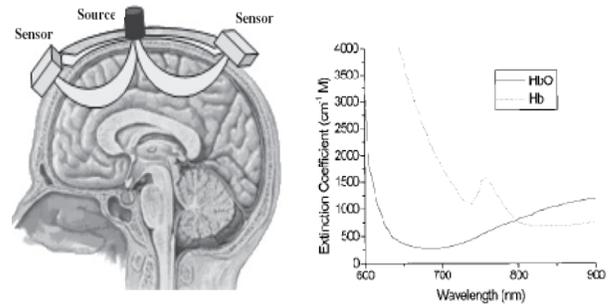

Figure 1: Light Propagation Path inside the Skull [24] & Absorption Spectra of HHb and HbO$_2$

NIRS can asses two types of hemodynamic changes associated with the brain activity. Increase in neural activity results in increased glucose and oxygen consumption, which leads to increase in HbO$_2$ concentration (*H. Matsuyama et al, 2009*).

Figure 3 shows HHb and HbO$_2$ signal acquisition while subject was doing a cognitive activity. It shows that each time when calculation was done, it caused a cognitive activation in the frontal region as demonstrated by increase in HbO$_2$ and decrease in HHb (*S. Perrey et al. 2010*).

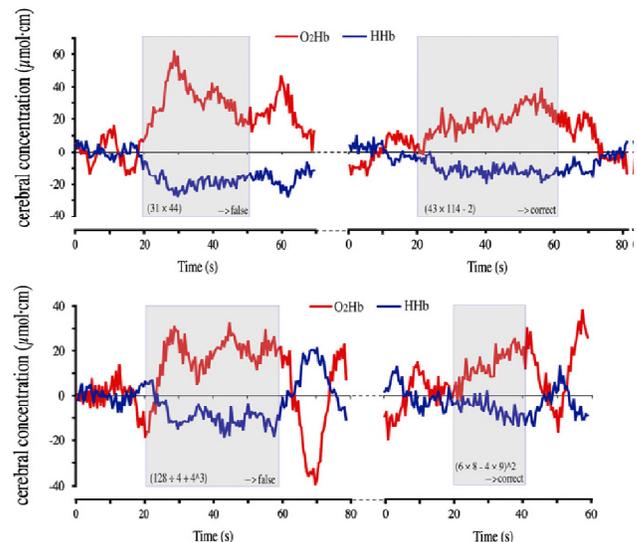

Figure 3: Calculation task was given to the subject, he was asked to resolve the arithmetic operation (as indicated in the gray box) with time pressure and precision demands [25]

### MEG: Magnetoencephalogram

Magnetoencephalography (MEG) is a noninvasive technique for investigating the magnetic field generated by the electrical activity of the neuronal population (*E.B.J. Coffey et al, 2010*). It records magnetic flux changes over the surface of the head (~10-15 Tesla) from synaptic discharge tangential currents due to the activity of neurons. MEG measurements are carried out in magnetically shielded rooms, using sensitive super-conducting quantum interference devices (SQUIDS) to detect these tiny magnetic fields. The MEG sensor consists of a flux transformer coupled to a SQUID, which considerably amplifies the weak extra cranial magnetic field and converts it into a voltage. It is possible to use MEG to study changes in brain activity even during high frequency deep brain stimulation. MEG data shares the basic features and frequency content of EEG, with predominant activity in delta band, frequency less than 4 Hz (*NJ Ray et al., 2007*).

## III. FUTURE OF CONCURRENT MEASUREMENT

In non-joint analysis we maximize the likelihood of functions for each modality separately, e.g. when we consider electrophiological response, hemodynamic response and brain activity separately. In contrast, for a joint analysis we join likelihood function, resulting in single fused unmixing parameter (*P. Fox et al, 1994*). Figure 4 is the Venn diagram showing the various possibilities for multimodal brain analysis.

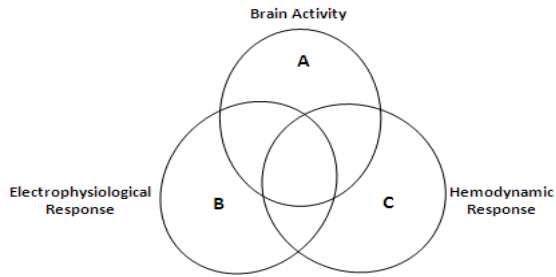

Figure 4: Venn diagram of Multimodal Analysis

Translating unimodal concepts from one modality to another is an important step towards understanding the neural mechanisms underlying these phenomena (*Mark E. Pflieger et al*). Neuronal decoding w.r.t any behavioral movement or cognitive movement can be correlated by going for a concurrent measurement of EEG, MEG and NIRS (*Y. O. Halchenko et al, 2005*). EEG, MEG and NIRS based non-invasive BMI development is designed with the objective of restoring the degree of mobility and communication in severely impaired patients who have lost all motor control because of spinal cord injury or who suffer from the locked in syndrome (*K. Jerbi et al., 2011*). Four multimodal paradigms are discussed below:

### 1. MEG and EEG

During concurrent measurement of EEG and MEG, the electrical signals measured from the surface of the head are correlated with the magnetic field generated by the motor cortex during the activity (*E.B.J. Coffey et al, 2010*).

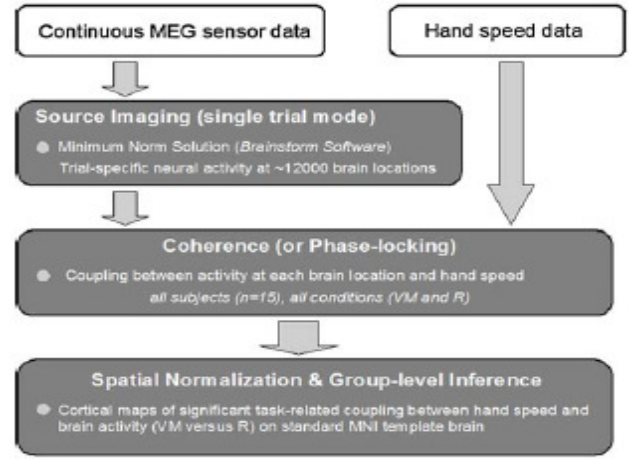

Figure 5: Illustrative Data Analysis Flow of MEG and EEG [8]

The figure 5 shows the illustrative data analysis flow for coherence between hand speed and neuromagnetic brain signals. In this case *K. Jerbi et al* has performed the data analysis with Brainstorm MEG and EEG toolbox.

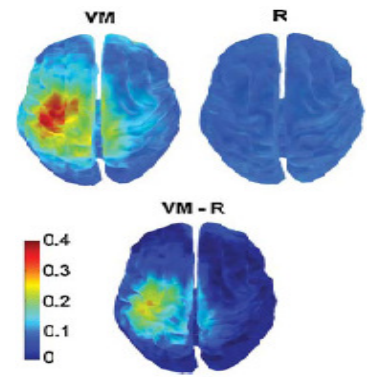

Figure 6: Coherence Maps between MEG and EEG [8]

Figure 6 shows the Z- Transformed coherence maps depicting low frequency coupling between cortical activity and time varying hand speed. The maps are shown for 'Visuomotor' – VM, the 'Rest Condition' – R and difference between the two (VM-R). The peak of the coherence between the brain and the hand speed was located in the contra-lateral primary motor cortex (*K. Jerbi et al., 2011*)

*Application:* This concurrent modality research offers the insight into non invasive brain computer interface (BCI) approach for the practical implementation. It is also used to measure additional information about epileptic activity, not seen when only EEG is measured.

## 2. EEG and fMRI

Electroencephalography and Functional Magnetic Resonance are two different methods for measuring neuronal activity in the brain. EEG provides excellent temporal resolution while fMRI preferred for its high spatial resolution *(D. Mantini et al, 2010.)* Concurrent analysis of EEG and fMRI is used to identify blood oxygen dependent (BOLD) changes associated with pshiological and pathological EEG events *(H. Laufs et al, 2003)*. Figure 7 shows the block diagram of concurrent analysis of EEG and fMRI. EEG was acquired simultaneously with fMRI by using 30 MR compatible electrodes with a sampling frequency of 5KHz. The main interest of this study is to create software which would combine EEG and fMRI to facilitate work of neurologist and researchers *(E. Martı́nez-Montes et al, 2004)*.

Relationship between EEG and fMRI is not precisely known, though few publications comment on negative correlation between alpha band of EEG and BOLD signal. *(R. Labounek et al, 2012)*. The finding suggests that power changes in EEG rhythms are associated with activity changes in the brain circuits *(H. Laufs et al, 2003)*.

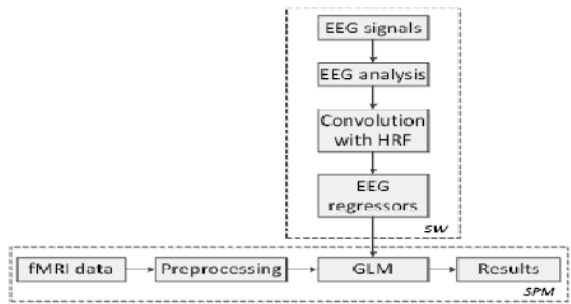

Figure 7: Block Diagram of Concurrent Analysis of EEG and fMRI [23]

*Application:* EEG-fMRI has potential to localize the neuronal activity with both high spatial and temporal resolution. This concurrent modality research offers the new possibilities in the investigations of brain rhythms, sleep patterns, and epilepsy. In the field of epilepsy, simultaneous EEG-fMRI is necessary for the study of the hemodynamic correlates of pathological discharges due to their subclinical nature. These studies have demonstrated BOLD increases and decreases in relation to sharp waves and sharp- and slow-wave complexes *(K. Blinowska et al, 2009)*.

## 3. NIRS and EEG

EEG-NIRS measurement depends on various physical properties such as conductivity, absorption and scattering coefficients of the head tissues such as scalp, skull, gray matter, white matter and cerebral blood flow (CBF).

NIRS requires the light in near infrared (NIR) region to determine cerebral oxygenation, blood flow and metabolic status of the brain. It provides non-invasive means of monitoring the brain function and biological tissue because of relatively low absorption by water and high absorption by HHb and HbO2 in the range of 600-1000 nm wavelength *(Herve´ F et al, 2008)*.

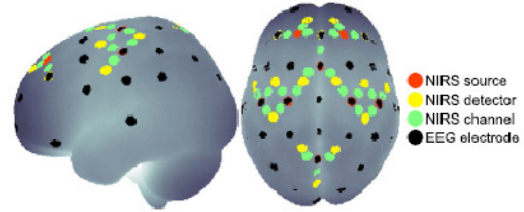

Figure 8: Locations of EEG Electrodes and Source and Detectors of NIRS System [26]

Figure 8 shows the placement of EEG electrodes along with the NIRS source and detectors for the EEG-NIRS concurrent analysis *(S. Fazli et al, 2012)*.

*Application:* This concurrent modality has been used to investigate the synchronized activities of neurons and the subsequent hemodynamic response in human subjects. This simple and comparatively low-cost setup allows to measure hemodynamic activity in many situations when fMRI measurements are not feasible, e.g. for long-term monitoring at the bedside or even outside the lab via wireless transmission.

## 4. fMRI and NIRS

NIRS signals correlate highly with BOLD fMRI. The strong correlation between the two means that many fMRI findings of regional activity specificity in the cerebral cortex can be used to guide NIRS research applications, and to better understand experimental results *(E.B.J. Coffey et al, 2010)*.

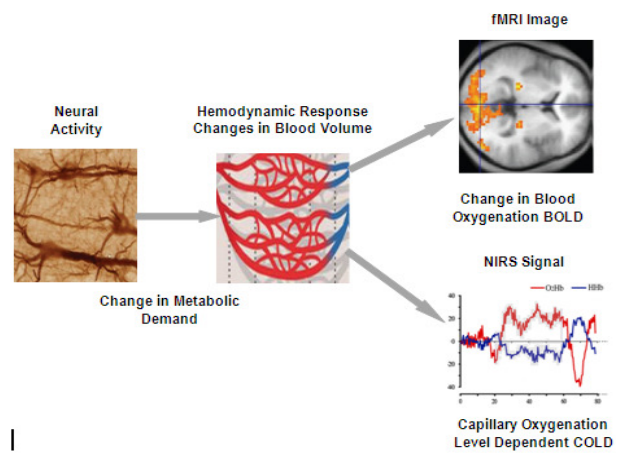

Figure 9: Neuronal Correlates of BOLD & COLD Signal [27]

Figure 9 shows the chain of events and factors that link neural activities to BOLD signal in fMRI image and COLD in NIRS signal. Neural activity through neurovascular coupling influences the metabolic demand. Metabolic changes impact on hemodynamic response which is dependent on physiological factors such as local cerebral blood flow, HHb/HbO$_2$ ratio, blood volume, and vascular geometry (*K. Blinowska et al, 2009)*. When a brain area is activated, metabolic activity increases, leading to a brief decrease in HbO$_2$ and increase in HHb about 2 s in the immediate vicinity of the activated neurons. This stimulates the increase of blood flow to a wider area, which causes HbO$_2$ levels to begin to increase to a peak at about 5s following neural firing, and then slowly declining over about 5–10 s after neural activity returns to normal (*E.B.J. Coffey et al, 2010)*.

<u>Application:</u> This concurrent modality research offers the better understanding of brain activation w.r.t. cognitive and behavioral changes.

## IV. Comparision of Various Modalities

Various brain modalities are compared in table 1, based on spatial resolution, temporal resolution. Advantages, disadvantages and applications of each modality are listed in the same table.

Table 1: Overview of Neuroimaging Modalities

| Imaging Method | Resolution | Application | Advantages | Disadvantages |
|---|---|---|---|---|
| *Functional Imaging Methods* | | | | |
| EEG | S - Low T - High | Study various rhythms, epilepsy, preoperative mapping, degenerative disorders | Non-invasive, no ionizing radiation, widely used, low cost | Low spatial resolution |
| MEG | S - Medium T - High | Study epilepsy | Non-invasive, no ionizing radiation, can identify epileptic foci | Low spatial resolution |
| fMRI | S - High T - Low | Preoperative mapping, functional mapping | Non-invasive, can perform functional imaging | High cost |
| NIRS | S - Low T - High | Functional mapping | Non-invasive, low cost, no ionizing radiation | Low spatial resolution |

S - Spatial Resolution; T - Temporal Resolution

## V. Conclusion

MEG and EEG provide an excellent temporal resolution of neuronal dynamics; while fMRI provide an alternative measure of neural activation based on hemodynamic changes in the brain with a very good spatial resolution. Near-infrared spectroscopy (NIRS) is a non-invasive method that enables real-time monitoring of tissue oxygenation of the brain because of this it is receiving increasing interest as a functional neuroscientific technique, complementing neuroelectric approaches such as EEG. NIRS can be applied in a variety of conditions as bedside monitoring in intensive care and in the operating theatre, where fMRI can be difficult to apply. All these non-invasive brain modalities complement and restrain each other and hence improve our understating of functional and neuronal organization. Spatially, temporally, physiologically, behaviourly and cognitively accurate computational models of the neuronal systems are the ultimate goals of the functional brain imaging. This goal can be achieved by integrating the diversity of various brain mapping techniques. By combining the various modalities together, we can exploit the strengths and flaws of individual brain imaging methods.

## Acknowledgment

The authors are grateful to Dr. Madhuri Khambete for her help towards the completion of this paper, as well as for providing valuable advice.

I would also like to thank my colleagues from Instrumentation and Control Dept. of Cummins College of Engineering for Women for their feedback during the discussions.

## References

[1] Gary Strangman, David A. Boas, and Jeffrey P. Sutton, "*Non-Invasive Neuroimaging Using Near-Infrared Light*", Society of Biological Psychiatry, Vol 52:679–693, 2005.

[2] A.W. Toga, P.M. Thompson, "*The Role of Image Registration in Brain Mapping*", Image and Vision Computing 19 (2001) 3–24.

[3] D. Salas-Gonzalez, J.M. Górriza, J. Ramírez, I. Álvarez, M. López, F. Segovia, C.G. Puntonet, "*Two Approaches to Selecting Set of Voxels for the Diagnosis of Alzheimer disease using Brain SPECT Images*", Digital Signal Processing 21 (2011) 746–755.

[4] T.S. Koh, Z. Hou, "*A Numerical Method for Estimating Blood Flow by Dynaic Functional Imaging*", Medical Engineering & Physics 24 (2002) 151–158.

[5] Ripen Misri, Dominik Meier, Andrew C. Yung, Piotr Kozlowski, Urs O. Häfeli, "*Development and Evaluation of a Dual-Modality (MRI/SPECT) Molecular Imaging Bioprobe*", Nanomedicine: Nanotechnology, Biology, and Medicine xx (2011) xxx–xxx (Article is in Press).

[6]    Augusto Buchweitz, Robert A. Mason, Lêda M. B. Tomitch and Marcel Adam Just, "*Brain Activation for Reading and Listening Comprehension: An fMRI Study of Modality Effects and Individual Differences in Language Comprehension*", Psychology & Neuroscience, 2009, 2, 2, 111 – 123.

[7]    Peter T Fox, Marty G Woldorff, "*Integrating Human Brain Maps*", Current Opinion in Neurobiology 1994, 4:151-156.

[8]    K. Jerbi, J.R. Vidal, J. Mattout, E. Maby, F. Lecaignard, T. Ossandon, C.M. Hamamé, S.S. Dalal, R. Bouet, J.-P. Lachaux, R.M. Leahy, S. Baillet, L. Garnero, C. Delpuech, O. Bertrand, "*Inferring hand movement kinematics from MEG, EEG and intracranial EEG: From brain-machine interfaces to motor rehabilitation*", IRBM 32 (2011) 8–18.

[9]    H. Laufs, A. Kleinschmidt, A. Beyerle, E. Eger, A. Salek-Haddadi, C. Preibisch, K. Krakow, "*EEG-correlated fMRI of Human Alpha Activity*", H. Laufs et al. / NeuroImage 19 (2003) 1463–1476

[10]   Sarah J. Erickson, Anuradha Godavarty, "*Hand-held Based Near-infrared Optical Imaging Devices: A Review*", Medical Engineering & Physics 31 (2009) 495–509.

[11]   H. Matsuyama, H. Asama, and M. Otake, "*Design of differential Near-Infrared Spectroscopy based Brain Machine Interface*", The 18th IEEE International Symposium on Robot and Human Interactive Communication Toyama, Japan, Sept. 27-Oct. 2, 2009.

[12]   Mamoru Tamura, Yoko Hoshi" Fumihiko Okada, "*Localized Near-Infrared Spectroscopy and Functional Optical Imaging Of Brain Activity*", Phil. Trans. R. Soc. Lond. B (1997) 352, 737±742, # 1997 The Royal Society, Printed in Great Britain.

[13]   Farzin Irani, Steven M. Platek, Scott Bunce, Anthony C. Ruocco, Douglas Chute, "*Functional Near Infrared Spectroscopy (fNIRS): An Emerging Neuroimaging Technology with Important Applications for the Study of Brain Disorders*", The Clinical Neuropsychologist, Volume 21, Issue 1 January 2007, pages 9 – 37.

[14]   Tomi Heinonen, Antti Lahtinen, Veikko Hakkinen, "*Implementation of Three-Dimensional EEG Brain Mapping*", Computers and Biomedical Research 32, 123–131 (1999).

[15]   K. Jerbi, J.R. Vidal, J. Mattout, E. Maby, F. Lecaignard, T. Ossandon, C.M. Hamamé, S.S. Dalal, R. Bouet, J.-P. Lachaux, R.M. Leahy, S. Baillet, L. Garnero, C. Delpuech, O. Bertrand, "*Inferring hand movement kinematics from MEG, EEG and intracranial EEG: From brain-machine interfaces to motor rehabilitation*", IRBM 32 (2011) 8–18.

[16]   Yaroslav O. Halchenko, Stephen Jos´e Hanson, Barak A. Pearlmutter, "*Multimodal Integration: fMRI, MRI, EEG, MEG*", Appears as pages 223-265 of Advanced Image Processing in Magnetic Resonance Imaging, Dekker, book series on Signal Processing and Communications, ISBN 0824725425, 2005.

[17]   Mark E. Pflieger, Randall L. Barbour, "*Multimodal Integration of fMRI, EEG, and NIRS*".

[18]   D. Mantini, L. Marzetti, M. Corbetta, G. L. Romani, C. Del Gratta, "*Multimodal Integration of fMRI and EEG Data for High Spatial and Temporal Resolution Analysis of Brain Networks*", Brain Topogr (2010) 23:150–158.

[19]   Eduardo Martı´nez-Montes, Pedro A. Valde´s-Sosa, Fumikazu Miwakeichi, Robin I. Goldman, Mark S. Cohen, "*Concurrent EEG/fMRI analysis by multiway Partial Least Squares*", NeuroImage 22 (2004) 1023–1034.

[20]   N. Nagdyman, T. P. K. Fleck, P. Ewert, H. Abdul-Khaliq, M. Redlin, P. E. Lange, "*Cerebral oxygenation measured by near-infrared spectroscopy during circulatory arrest and cardiopulmonary resuscitation*", British Journal of Anaesthesia 91 (3): 438±42 (2003).

[21]   NJ Ray, ML Kringelbach, N Jenkinson, SLF Owen, P Davies, S Wang, N De Pennington, PC Hansen, J Stein, TZ Aziz, "*Using magnetoencephalography to investigate brain activity during high frequency deep brain stimulation in a cluster headache patient*", Biomedical Imaging and Intervention Journal, doi: 10.2349/biij.3.1.e25

[22]   R. Labounek, M. Lamoš, R. Mareček, J. Jan, "*Analysis of Connections between Simultaneous EEG and fMRI Data*", IWSSIP 2012, 11-13 April 2012, Vienna, Austria, ISBN 978-3-200-02328-4.

[23]   Herve´ F. Achigui, Mohamad Sawan, Christian J.B. Fayomi, "*A monolithic Based NIRS Front-end Wireless Sensor*", Microelectronics Journal 39 (2008) 1209–1217.

[24]   Emily B.J.Coffey, Anne-MarieBrouwer, EllenS.Wilschut, JanB.F.vanErp, "*Brain–Machine Interfaces In Space: Using Spontaneous Rather Than Intentionally Generated Brain Signals*", Acta Astronautica 67 (2010) 1–11, doi:10.1016/j.actaastro.2009.12.016

[25] Stephane Perrey, Thibaud Thedon, Thomas Rupp, "*NIRS in ergonomics: Its application in industry for promotion of health and human performance at work*", International Journal of Industrial Ergonomics 40 (2010) 185–189.

[26] Siamac Fazli, Jan Mehnert, Jens Steinbrink, Gabriel Curio, Arno Villringer, Klaus-Robert Müller, Benjamin Blankertz, "*Enhanced Performance By A Hybrid NIRS–EEG Brain Computer Interface*", NeuroImage 59 (2012) 519–529

[27] Katarzyna Blinowska, GernotM¨uller-Putz, Vera Kaiser, Laura Astolfi, Katrien Vanderperren, Sabine Van Huffel, Louis Lemieux, "*Multimodal Imaging of Human Brain Activity: Rational, Biophysical Aspects and Modes of Integration*", Computational Intelligence and Neuroscience Volume 2009, doi:10.1155/2009/813607.

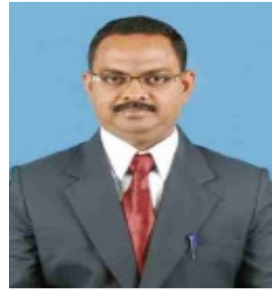

**Dr. M. Sundhararajan** received MS degree from Birla Institute of Technology & Science (BITS) Pilani and PhD from Bharathidasan University, Trichy, INDIA. He is currently working as a Principal in Shri Laxmi Ammal College of Engineering, Chennai, INDIA

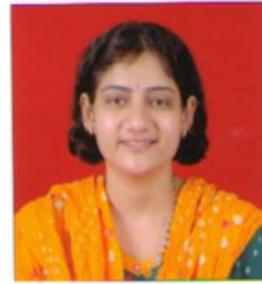

**Nivedita Daimiwal** received the B.E. and M.E. degree in Biomedical Instrumentation from University of Pune. She is currently working towards the Ph.D. degree at Sathyabama University, Chennai. She is currently working as an Assistant Professor in MKSSS's Cummins College of Engineering for women, Pune, INDIA.

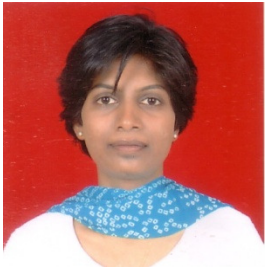

**Revati Shriram** received the B.E. degree in Instrumentation and Control from University of Pune, M.S. in Electrical Engineering from Rose-Hulman Institute of Technology, Indiana, USA. She is currently working towards the Ph.D. degree at Sathyabama University, Chennai. She is currently working as an Assistant Professor in MKSSS's Cummins College of Engineering for Women, Pune, INDIA.